\documentclass[conference]{IEEEtran}
\IEEEoverridecommandlockouts
\usepackage{cite}
\usepackage{amsmath,amssymb,amsfonts}
\usepackage{algorithmic}
\usepackage{graphicx}
\usepackage{textcomp}
\usepackage{xcolor}
\usepackage{graphicx}
\usepackage{subfigure}
\usepackage{multicol}
\usepackage{setspace}
\usepackage{amsmath}
\usepackage{epstopdf}
\usepackage{fancyhdr}
\usepackage{indentfirst}
\usepackage{enumerate}
\usepackage{caption}
\usepackage{leftidx}
\usepackage{amssymb}
\usepackage{float}
\usepackage{breqn}
\usepackage{bbm}
\usepackage{booktabs}
\usepackage{bm}
\usepackage{multirow}
\usepackage{totcount}
\usepackage{algorithm}
\allowdisplaybreaks[4]
\def\BibTeX{{\rm B\kern-.05em{\sc i\kern-.025em b}\kern-.08em
		T\kern-.1667em\lower.7ex\hbox{E}\kern-.125emX}}

\begin{document}
	
	\title{Joint Device Identification, Channel Estimation, and Signal Detection for LEO Satellite-Enabled \\Random Access}
	
	\author{Boxiao Shen, Yongpeng Wu, Wenjun Zhang, Symeon Chatzinotas, and Björn Ottersten 
		\thanks{B. Shen, Y. Wu, and W. Zhang are with the Department of Electronic Engineering, Shanghai Jiao Tong
			University, Shanghai 200240, China (e-mails:
			\{boxiao.shen, yongpeng.wu, zhangwenjun\}@sjtu.edu.cn).
		}
		\thanks{
			Symeon Chatzinotas and Björn Ottersten are with the Interdisciplinary Center for Security, Reliability
			and Trust (SnT), University of Luxembourg, 1855 Luxembourg City, Luxembourg (e-mails: 
			\{Symeon.Chatzinotas, bjorn.ottersten\}@uni.lu).
		}
	}
	
	\maketitle
	
	\begin{abstract}
		This paper investigates joint device identification, channel estimation, and signal detection for LEO satellite-enabled grant-free random access, where a multiple-input multiple-output (MIMO) system with orthogonal time-frequency space modulation (OTFS) is utilized to combat the dynamics of the terrestrial-satellite link (TSL).
		We divide the receiver structure into three modules: 
		first, a linear module for identifying active devices, which leverages the generalized approximate message passing (GAMP) algorithm to eliminate inter-user interference in the delay-Doppler domain; second, a non-linear module adopting the message passing algorithm to jointly estimate channel and detect transmit signals; the third aided by Markov random field (MRF) aims to explore the three dimensional block sparsity of channel in the delay-Doppler-angle domain. The soft information is exchanged iteratively between these three modules by careful scheduling. Furthermore, the expectation-maximization algorithm is embedded to learn the hyperparameters in prior distributions. Simulation results demonstrate that the proposed scheme outperforms the conventional methods significantly in terms of activity error rate, channel estimation accuracy, and symbol error rate. 
	\end{abstract}
	
	
	\section{Introduction}
	 Internet-of-Things (IoT) is one of the critical scenarios in next-generation communications. Presently, there are numerous applications where IoT devices could be distributed in remote regions, such as deserts, oceans, and forests \cite{s41}, but they are not supported by existing cellular communication networks. Fortunately, low-earth orbit (LEO) satellites possess low propagation delay, low path loss, and flexible elevation angles, making them a highly promising solution to provide global coverage. In such cases, the multiple access protocol plays a key role in supporting efficient connectivity.
	
	Grant-free random access (GFRA) is considered to be suitable for machine-type communications, as it reduces signaling overhead and power consumption, and enhances access capability. Over the past few years, many methods have been proposed for joint channel estimation and device identification in the terrestrial GFRA systems. For example, the approximate message passing (AMP) was adopted in \cite{s33} to address this problem. In \cite{s43}, orthogonal
	frequency division multiplexing (OFDM) was integrated into GFRA systems, and the generalized multiple measurement vector AMP was proposed to explore the channel sparsity in the angular domain. To further improve system performance, \cite{Joint1} and \cite{Joint2} adopted spreading-based transmission schemes and designed message passing-type algorithm to jointly identify devices, estimate channel, and detect signal. The above mentioned works have been designed for block fading channels, which are assumed to remain constant during once transmission. However, the high mobility of LEO satellite inevitably leads to rapid change of terrestrial-satellite link (TSL) and the large Doppler shift, and the motion of devices may incur nonnegligible Doppler spread \cite{ss32}, both of which may cause outdated CSI and severe inter-carrier interference, and then degrades the performance of current algorithms. These effects and the propagation delay should be taken into account. As a result, current terrestrial GFRA schemes are not directly applicable to LEO satellite communications. To facilitate transmissions, precompensation technique \cite{cfocompen} for delay and Doppler shift may be adopted before GFRA, but it introduces extra complexity which may decrease the battery life of remote IoT devices, and the most of conventional schemes cannot handle the Doppler spread effectively.
	
	Orthogonal time-frequency space (OTFS) modulation \cite{s9},  which operates directly in the delay-Doppler domain, has been proposed as a promising solution to alleviate the aforementioned issues. The location of nonzero element of effective channel corresponds to the delay and Doppler shift of each physical path when the two conditions are satisfied that the delay and Doppler shift are within one symbol duration and subcarrier spacing, respectively. Therefore, it is possible for satellite to estimate the effective channel and then detect signal, without precompensation at the terrestrial devices.
	In \cite{smm1} and \cite{smm2}, two GFRA schemes with MIMO-OTFS have been proposed for LEO satellite communications, where the channel estimation and signal detection are considered separately. However, to facilitate algorithm design, they still require the terrestrial devices to precompensate for delay and/or Doppler shift before GFRA.
	
	In this paper, we investigate joint device identification, channel estimation, and signal detection in LEO satellite-enabled GFRA, where MIMO-OTFS is adopted to address the doubly dispersive effect and improve performance. Different from the previous literature, we assume that the IoT devices lack global navigation satellite system (GNSS) capability and do not need to precompensate for delay and Doppler shift. The satellite will handle them in the uplink transmission, and thus the complexity and energy consumption of terrestrial devices are reduced. In this scenario, the propagation delay will be more than one symbol duration and/or the Doppler shift will be more than one subcarrier spacing, which brings the extra phase rotation into the effective channel. As a result, the effective channel at each antenna is respresented as a three dimensional (3D) tensor, and simultaneously, is coupled together with the device activities and transmit signals, forming a complicated non-linear signal model. Furthermore, the channel tensor exhibits 3D block sparsity in the delay-Doppler-angle (DDA) domain. 
	By introducing appropriate auxiliary variables, we divide the whole detection scheme into three modules to address this problem: delay-wise device activity detection (DDAI), combined channel estimation and signal detection (CCESD), and 3D sparsity exploration (TSE). Specifically, in the DDAI, we handle the received signal along the delay dimension such that the 3D channel tensor can be split as a series of matrices, and then the generalized AMP (GAMP) algorithm is adopted to decouple the transmissions of different devices in different delay-Doppler dimensions. Otherwise, the severe inter-user and inter-carrier interference will deteriorate the transmission quality. The CCESD module deals with the nonlinear coupling of the activity state, channel coefficient, and transmit signals of each device, where a message passing algorithm is
	derived in a symbol-by-symbol fashion for each delay dimension. The TSE module adopts the Markov random field (MRF) to capture the 3D block sparsity of the channel tensor. By carefully message scheduling, the three modules exchange the soft information with each other iteratively until convergence. Furthermore, the expectation-maximization (EM) algorithm is embedded to learn the hyperparameters in priors.
	
	\section{System Model}
	
	We consider a LEO satellite-enabled GFRA system with MIMO-OTFS. The system involves $U$ single-antenna devices that aim to communicate with a LEO satellite, which is equipped with a uniform planar array (UPA) of $N_a = N_y \times N_z$ antennas and a regenerative payload capable of on-board processing of baseband signals. In each time interval, each device shares the same time-frequency resources to transmit signal to the satellite with probability $p_{\lambda}$. In addition, we consider the scenario that the ground devices lack GNSS capability. Following the recommendations of the 3GPP \cite{3gpp}, in this scenario, the satellite will firstly precompensate a common delay to all devices, and then handle the differential delay and Doppler shift seen in the uplink transmission. In the next subsections, we first introduce the input-output relationship of the system, and then formulate the considered problem.
		
	\subsection{Input-Output Relationship}
	Due to the rapid variations of TSL, we consider the doubly dispersive channel \cite{g1} in this work. Then, we adopt the spreading-based scheme \cite{Joint1} \cite{Joint2} for grant-free transmission and the OFDM-based OTFS modulation to combat the doubly dispersive effect of the TSL. Specifically, in the $q$-th OTFS frame, $q=0,\dots,Q-1$, the $u$-th device is assigned with a unique spreading code $\mathbf C_u^q[k,l]$, $k=\lceil -N/2\rceil,\dots,\lceil N/2\rceil-1$, $l=0,\dots,M-1$, where $M$ and $N$ are the number of subcarriers and OFDM symbols within one OTFS frame, respectively. Then, the information symbol $\mathbf t_u[l]$ is spread into $Q$ frames, i.e., the transmitted signal in the delay-Doppler domain is $\mathbf C_u^q[k,l] \mathbf t_u[l]$, where $\mathbf t_u[l]$ is selected from a predefined alphabet $\mathcal{A} = \{a_1, \dots, a_{\left|\mathcal{A}\right|}\}$ with cardinality $\left|\mathcal{A}\right|$. Next, the transmitted signal will go through the OTFS modulation, doubly dispersive channel, and OTFS demodulation. The detailed process is omitted here due to the limited spacing and interested readers can refer to \cite{s22,smm1}. To handle the large differential delay, we require the cyclic prefix (CP) duration adopted by each device to be greater than it. Then, based on the results in \cite{smm1}, the $q$-th received frame (without noise) of the $u$-th device in the delay-Doppler-angle domain is given by
	\begin{align}
	\label{ryDDA}
	\mathbf Y_{u,a_y, a_z}^{\mathrm{DDA}^q}[k, l] =\sum_{l^{\prime}=0}^{M-1} \sum_{k^{\prime}=\lceil-N / 2\rceil}^{\lceil N / 2\rceil-1} 
	\mathbf H_{u, a_y,a_z}^{\mathrm{DDA}}\left[k^{\prime}, l^{\prime}, l\right] \nonumber \\
	\times \mathbf t_u\left[\left(l-l^{\prime}\right)_{M}\right] 
	\mathbf C_u^q\left[\left\langle k-k^{\prime}\right\rangle_{N},\left(l-l^{\prime}\right)_{M}\right],
	\end{align}   
	where $a_y = 0,\dots,N_y-1$ and $a_z = 0,\dots,N_z-1$ are indexes in the angular domain, $(\cdot)_{M}$ denotes mod $M$, $\langle x\rangle_{N}$ denotes $\left(x+\left\lfloor\frac{N}{2}\right\rfloor\right)_{N}-\left\lfloor\frac{N}{2}\right\rfloor$, and $\mathbf H_{u,a_y,a_z}^{\mathrm{DDA}}$ is the effective channel in the delay-Doppler-angle domain, represented as
    \begin{align}
    	\label{angleD}
    	&\mathbf H_{u, a_y,a_z}^{\mathrm{DDA}}[k^{\prime}, l^{\prime}, l] 
    	=\sqrt{N_yN_z}\sum_{i=1}^{P}h_{u,i} e^{\bar{\jmath} 2 \pi(M_{\text{cp}} + l)T_s\nu_{u,i}} \nonumber \\
    	&\times e^{-\bar{\jmath} 2 \pi\tau_{u,i}\nu_{u,i}} \Pi_N(k^{\prime}-NT_{\text{sym}}\nu_{u,i})\delta\left(l^{\prime} T_{\mathrm{s}}-(\tau_{u,i})_T\right)\nonumber\\  &\times\Pi_{N_y}(a_y-N_y\vartheta_{y_{u,i}}/2) \Pi_{N_z}(a_z-N_z\vartheta_{z_{u,i}}/2),
    \end{align}
    where $h_{u,i}$,$\tau_{u,i}$, and $\nu_{u,i}$ are the complex gain, differential delay, and Doppler shift, respectively; $\vartheta_{y_{u,i}}$ and $\vartheta_{z_{u,i}}$ are the directional cosines along the $y$- and $z$-axis of UPA, respectively; $M_{\text{cp}}$ is the length of CP, $T$ is one symbol duration, $T_{\text{sym}}$ is $T$ plus CP duration, $T_s$ is the system sample rate, and $\Pi_N(x)\triangleq\frac{1}{N} \sum_{i=0}^{N-1} e^{-\bar{\jmath} 2 \pi \frac{x}{N} i}$. From (\ref{angleD}), $\mathbf H_{u, a_y,a_z}^{\mathrm{DDA}}[k^{\prime}, l^{\prime}, l]$ has dominant elements only if $k^{\prime} \approx NT_{\text{sym}}(\nu_{u,i})_{1/T_{\text{sym}}}$, $l^{\prime} \approx (\tau_{u,i})_T/T_s$, $a_y \approx N_y\vartheta_{y_{u,i}}/2$, and $a_z \approx N_z\vartheta_{z_{u,i}}/2$. Therefore, the channel in the delay-Doppler-angle domain shows the 3D-structured sparsity\cite{smm1}. In addition, unlike that in previous literature\cite{s22}, the effective channel has an extra dimension related to the delay dimension $l$ of the received signal, which results in the 3D channel tensor at each antenna,
    since in the LEO satellite communications, the large differential delay and Doppler shift usually cannot meet the conditions, $\tau_{u,i} < T$ and $\nu_{u,i} < \Delta f$, simultaneously \cite{3gpp}. 
     Hence, the efficient algorithm is necessary to be designed for this situation.
    \vspace{-0.1cm}
	\subsection{Problem Formulation}
	The $q$-th frame received from all devices can be represented as
	\begin{align}
		\label{allsig}
		\mathbf Y_{q,a_y, a_z}^{\mathrm{DDA}}[k, l] = \sum_{u=0}^{U-1} \lambda_u \mathbf Y_{u,a_y, a_z}^{\mathrm{DDA}^q}[k, l]+ \mathbf Z_{q,a_y, a_z}^{\mathrm{DDA}}[k, l],
	\end{align}
	where $\lambda_u$ is the activity indicator of the $u$-th device, with $\lambda_u = 1$ if active and $\lambda_u=0$ otherwise, and $\mathbf Z_{q,n_y, n_z}^{\mathrm{DDA}}[k, l] \sim \mathcal{CN}(0,\sigma^2)$ is the noise in the delay-Doppler-angle domain. Note that the effective channel of each device in (\ref{angleD}) is a 3D tensor at each antenna.  To facilitate the algorithm design, we decouple it into a set of matrices by rewriting (\ref{allsig}) as the matrix form along the received delay dimension $l$, i.e.,
	\begin{align}
	\label{qframe}
	\mathbf Y_q^l =  \mathbf C^l_q (\mathbf T^l \otimes \mathbf I_N )(\Lambda \otimes \mathbf{I}_{MN}) \tilde{\mathbf H}^l + \mathbf Z_q^l,
	\end{align}
	where the $(a_z+1+N_za_y)$-th column of $\mathbf Y^l_q$ is $\mathbf Y_{q,a_y, a_z}^{\mathrm{DDA}}[:,l] \in\mathrm{C}^{N}$; 
	$\mathbf C^l_q=[\mathbf C_{0,q}^l,\dots,\mathbf C_{U-1,q}^l]$, where $\mathbf C_{u,q}^l\in\mathrm{C}^{N\times MN}$ is the spreading code matrix of the $u$-th device in received delay dimension $l$. Note that $\mathbf C_{u,q}^l$ is a block circulant matrix due to the 2D circular convolution in (\ref{ryDDA}), 
	and its sub-matrix is the circulant matrix, formed by $[\mathbf C_{u}^q[0,(l-l^{\prime})_M],\mathbf C_{u}^q[1,(l-l^{\prime})_M],\dots,\mathbf C_{u}^q[-1,(l-l^{\prime})_M]]$. $\mathbf T^l = \mathrm{diag}([\mathbf t_0^l, \dots, \mathbf t_{U-1}^l])$, where $\mathrm{diag}(\mathbf x)$ returns a diagonal matrix with the elements of vector $\mathbf x$ on the main diagonal, $\mathbf t_u^l=[\mathbf t_u[(l)_M],\dots,\mathbf t_u[(l-M+1)_M]]$, and $\mathbf{I}_{N} \in \mathrm{R}^{N\times N}$ is the identity matrix;
	$\Lambda = \mathrm{diag}(\bm{\lambda})$, where $\bm{\lambda} = [\lambda_0,\cdots,\lambda_{U-1}]$; the $(a_z+1+N_za_y)$-th column of $\tilde{\mathbf H} \in \mathrm{C}^{UMN\times N_a}$ is given by $\left[\mathrm{vec}^{\text T}\left(\mathbf H^{\mathrm{DDA}}_{0,a_y,a_z}[:,:,l]\right),\dots,\mathrm{vec}^{\text T}\left(\mathbf H^{\mathrm{DDA}}_{U-1,a_y,a_z}[:,:,l]\right) \right]^{\text T} \in \mathrm{C}^{UMN}$, where $\mathrm{vec}(\cdot)$ denotes the vectorization of a matrix; the elements of $\mathbf Z^l_q$ are independent Gaussian noises. 
	We then collect all the received frames into a single matrix as 
	\begin{align}
		\label{allframe}
		\mathbf Y^l &=  \mathbf C^l (\mathbf T^l \otimes \mathbf I_N ) \mathbf H^l + \mathbf Z^l \\ 
		&= \mathbf C^l \mathbf W^l + \mathbf Z^l, \label{wr}
	\end{align}
	where $\mathbf Y^l = \left[(\mathbf Y_0^l)^{\text T},\dots,(\mathbf Y_{Q-1}^l)^{\text T}\right]^{\text T}$,  $\mathbf C^l = \left[(\mathbf C_0^l)^{\text T},\dots,(\mathbf C_{Q-1}^l)^{\text T}\right]^{\text T}$, $\mathbf W^l = (\mathbf T^l \otimes \mathbf I_N ) \mathbf H^l$, and $\mathbf H^l = (\Lambda \otimes \mathbf{I}_{MN})  \tilde{\mathbf H}^l$. 
	
	Next, to clearly elaborate the formulated problem and the proposed algorithm, we introduce some notations. Firstly, $\mathbf H^l$ is partitioned as submatrices $\mathbf H^{l,u}=\mathbf H^l[uMN:(u+1)MN-1,:] \in \mathrm{C}^{MN \times N_a}$ corresponding to the channel matrix of the $u$-th device along the received delay dimension $l$. Those submarices are further split as $\mathbf H^{l,u,l^{\prime}} = \mathbf H^{l,u}[l^{\prime}N:(l^{\prime}+1)N-1,:] \in \mathrm{C}^{N \times Na}$, $l^{\prime}=0,\dots,M-1$, which is the channel of the $l^{\prime}$-th grid in the delay dimension of the $u$-th device. $\mathbf W^l$ is partitioned similarly, and then we get $\mathbf W^{l,u}$ and $\mathbf W^{l,u,l^{\prime}}$.  Finally, we collect all the transmitted information symbols as the vector $\mathbf t = [\mathbf t_0,\dots,\mathbf t_{U-1}]$, and collect all the channels as $\mathbf H = [\mathbf H^0,\dots,\mathbf H^{M-1}]$. 
	
    To perform joint device identification, channel estimation and signal detection, we resort to the Bayesian approach which needs prior distribution of the estimated variables. Firstly, we adopt the conditional Bernoulli Gaussian-mixture (GM) distribution to characterize the channel, i.e.,
    \begin{align}
    	\label{hprior}
    	p\left(h_{i, j}^{l, u, l^{\prime}} \mid s_{i, j}^{u, l^{\prime}} \right) = 
    	&\delta\left(s_{i, j}^{u, l^{\prime}}-1\right) \sum_{k=1}^K \omega_k^u \mathcal{CN}\left(h_{i, j}^{l, u, l^{\prime}}\mid \mu^u_k, \phi^u_k\right) \nonumber \\
    	&+ \delta\left(s_{i, j}^{u, l^{\prime}}+1\right) \delta\left(h_{i, j}^{l, u, l^{\prime}}\right),
    \end{align}  
	where $K$ and $(\omega_k^u, \mu^u_k, \phi^u_k)$ are the number of components and parameters of GM, respectively, $h_{i, j}^{l, u, l^{\prime}}$ is the $(i,j)$-th element of $\mathbf H^{l,u,l^{\prime}}$, $s_{i, j}^{u, l^{\prime}} \in \{+1,-1\}$ is the corresponding support, and $\delta(\cdot)$ is the Dirac delta function. Then, we adopt the Markov random field (MRF) prior to describe the 3D block sparsity of the channel tensor, and then the support can be characterized by the classic Ising model as
	\begin{align}
	\label{sprior}
		p\left(\mathbf{S}^{u, l^{\prime}}\right) 
		&\propto \exp \left(\sum_{i=0}^{N-1} \sum_{j=0}^{N_a-1}\left(\frac{1}{2} \sum_{s_{i^{\prime}, j^{\prime}}^{u, l^{\prime}} \in \mathcal D_{i, j}^{u, l^{\prime}}} \beta s_{i^{\prime}, j^{\prime}}^{u, l^{\prime}}-\alpha\right) s_{i, j}^{u, l^{\prime}}\right)
		\nonumber \\
		&=\left[\prod_{i,j}\prod_{s_{i^{\prime}, j^{\prime}} \in \mathcal D_{i, j}}  
		\psi(s_{i,j}^{u,l^{\prime}}, s_{i^{\prime}, j^{\prime}}^{u, l^{\prime}})\right]^{\frac{1}{2}}
		\prod_{i,j}\gamma(s_{i,j}^{u,l^{\prime}})
	\end{align}
where $\psi(s_{i,j}^{u,l^{\prime}}, s_{i^{\prime}, j^{\prime}}^{u, l^{\prime}}) = \exp(\beta s_{i, j}^{u, l^{\prime}}
s_{i^{\prime}, j^{\prime}}^{u, l^{\prime}})$,
$\gamma(s_{i,j}^{u,l^{\prime}}) = \exp(-\alpha s_{i, j}^{u, l^{\prime}})$,
$\mathbf S^{u, l^{\prime}}$ is a support matrix with $(i,j)$-th element $s_{i, j}^{u, l^{\prime}}$, $\mathcal D_{i,j}^{u, l^{\prime}} = \{ s_{i-1,j}^{u,l^{\prime}}, s_{i+1,j}^{u,l^{\prime}}, s_{i,j-1}^{u,l^{\prime}}, s_{i,j+1}^{u,l^{\prime}} \}$ is the set containing the neighbors of $s_{i,j}^{u,l^{\prime}}$, and $\alpha$ and $\beta$ are the parameters of MRF prior; a larger $\beta$
implies a larger size of each block of nonzeros, and a larger $\alpha$ encourages a sparser $\mathbf H^{l,u,l^{\prime}}$.
\vspace{-0.3cm}
\begin{figure}[!htb]
	\centering
	\captionsetup{font={small}}
	\setlength{\belowcaptionskip}{-.3cm}
	\includegraphics[width=2.8in]{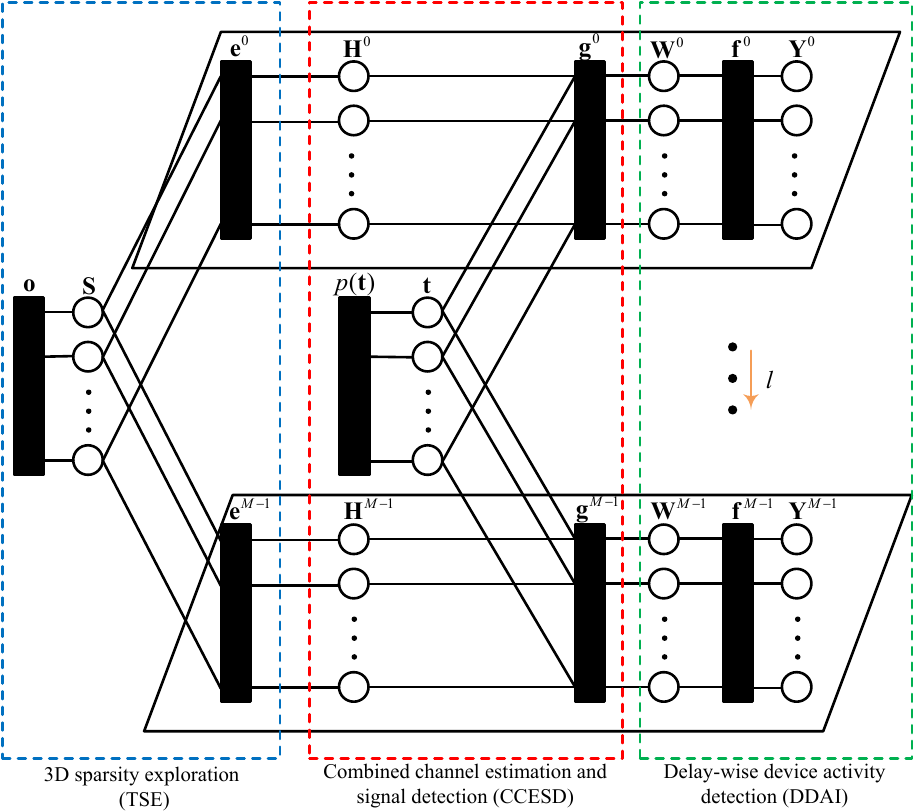}
	\caption{Factor graph representation.}
	\label{Factor}
\end{figure}
Based on (\ref{allframe})-(\ref{sprior}), the maximum a
posteriori (MAP) estimate for $(\mathbf W, \mathbf H, \mathbf t, \mathbf S)$ is given by
\begin{align}
\label{map}
(\hat{\mathbf W}, \hat{\mathbf H}, \hat{\mathbf t} , \hat{\mathbf S})=\underset{(\mathbf W, \mathbf H, \mathbf t, \mathbf S)}{\text{arg max}} p(\mathbf W, \mathbf H, \mathbf t, \mathbf S \mid \mathbf Y),
\end{align}
where $\mathbf Y = [\mathbf Y^0,\dots,\mathbf Y^{M-1}]$, $\mathbf W = [\mathbf W^0,\dots,\mathbf W^{M-1}]$, $\mathbf S$ is composed of $\mathbf S^{u,l^{\prime}}$, and the posterior distribution is represented as
\begin{align}
	\label{post}
	&p(\mathbf W, \mathbf H, \mathbf t, \mathbf S \mid \mathbf Y) \nonumber\\
	&\propto \prod_{l=0}^{M-1} p(\mathbf Y^l \mid \mathbf W^l) p(\mathbf W^l \mid \mathbf H^l, \mathbf t) p(\mathbf t) p(\mathbf H^l \mid \mathbf S) p(\mathbf S),
\end{align}
 where $p(\mathbf W^l \mid \mathbf H^l, \mathbf t)$ corresponds to the constraints in (\ref{allframe}), given as 
 \begin{align}
 	\label{constraint}
 	p(\mathbf W^l \mid \mathbf H^l, \mathbf t) = \delta(\mathbf W^l-(\mathbf T^l \otimes \mathbf I_N)  \mathbf H^l).
 \end{align}
 Then, the active devices can be detected by the energy detector, given by
 \begin{align}
 	\label{detectA}
 	\hat{\lambda}_u = \mathbb{I}\left\{\sum_{l=0}^{M-1} \Vert\mathbf W^{l,u}\Vert_F^2 > \xi_{th}\right\}, 
 \end{align}
 where $\mathbb{I}\{\cdot\}$ is the indicator function and $\xi_{th}$ is the predefined threshold. Note that problem (\ref{map}) is generally non-convex and difficult to solve.
 Since the variables to be estimated in (\ref{map}) are all coupled together, an accurate message passing algorithm design is challenging. We next develop a low-complexity iterative algorithm that could achieve near-optimal performance by utilizing a carefully designed receiver structure and sophisticated message updates. Additionally, the phase ambiguity problem \cite{Joint1} inevitably arises in (\ref{map}) since both channel and information symbols are unknown in the receiver. Common methods for combating this problem include differential coding or asymmetric constellation\cite{phase1}. In this work, we adopt the latter approach.

	\section{Joint Device Identification, Channel Estimation, and Signal Detection}
	In this section, we present the proposed algorithm for joint device identification, channel estimation and signal detection.
	Firstly, the factor graph is introduced for describing the probability structure defined in (\ref{post}). Then, based on the factor graph, the message passing-type method is designed for estimating the posterior distribution of variables, and the EM algorithm is embedded to learn the unknown hyperparameters in the prior distribution. 
	\subsection{Factor Graph Representation}
	The factor graph corresponding to (\ref{post}) is shown in Fig. \ref{Factor}, which consists of two types of nodes:
	\begin{itemize}
		\item Variable nodes $(\mathbf W, \mathbf H, \mathbf t, \mathbf S)$, depicted as white circles in Fig. \ref{Factor}, correspond to the variables with the same names in (\ref{allframe})-(\ref{post}).
		\item Check nodes $\{\mathbf f^l\}$, $\{\mathbf g^l\}$, $\{\mathbf e^l\}$, and $\{\mathbf o\}$ depicted as black rectangles in Fig. \ref{Factor}, correspond to the likelihood function of (\ref{wr}), constraint in (\ref{constraint}), the conditional probability density function (PDF) of $\mathbf H^l$, and the MRF prior of $\mathbf S$, respectively. The edge message passes between a variable node and a check
		node when the variable is involved in the check constraint.
	\end{itemize}
	In addition, as shown in Fig. \ref{Factor}, we divide the whole receiver structure into three modules:
	 DDAI module for the linear signal model in (\ref{wr}), aims to estimate $\mathbf W^l$ parallel along the received delay dimension $l$, which is adopted to identify active devices. Simultaneously, it decouples the transmissions in different delay-Doppler dimensions of different devices.
	 CCESD module handles the non-linear constraint in (\ref{constraint}), which takes the output soft messages of DDAI module and refined messages of TSE module as input, and then jointly performs channel estimation and signal detection.
	 TSE module for the MRF prior, considers the messages generated by CCESD module and explores the 3D block sparsity of $\mathbf H$. Then, the refined messages will be fed back to CCESD module.
The messages are passed between the three modules iteratively until convergence.
	\subsection{Posterior Distribution Estimation}
	In this subsection, with the known hyperparameters, we describe how the messages iterate between the three modules to get the final estimation, and the hyperparameters will be updated later. In the following content, we denote $\Delta^x_{x^{\prime}}$ as the message from node $x$ to $x^{\prime}$. Firstly, the DDAI module aims to estimate $\mathbf W$ in the linear model, and hence the output message can be approximated by the GAMP algorithm\cite{sss}, given as
	\begin{align}
	\label{gampout}
	\Delta_{g_{i,j}^{l,u,l^{\prime}}}^{w_{i,j}^{l,u,l^{\prime}}}
	=\mathcal{CN}(w_{i,j}^{l,u,l^{\prime}}\mid\hat{r}_{i,j}^{w^{l,u,l^{\prime}}}, \tau_{i,j}^{w^{l,u,l^{\prime}}}),
	\end{align}
 where the mean $\hat{r}_{i,j}^{w^{l,u,l^{\prime}}}$ and the variance $\tau_{i,j}^{w^{l,u,l^{\prime}}}$ are updated iteratively by GAMP algorithm. 
 Next, we focus on the messages scheduling of the CCESD and the TSE module.
	Combined the output of DDAI module with the message from variable nodes $\mathbf t$ to check nodes ${\mathbf g^l}$, the message from ${\mathbf g^l}$ to $\mathbf H$ will be a GM distribution given as
	\begin{align}
	\label{CCESDfirst}
		\Delta^{g_{i,j}^{l,u,l^{\prime}}}_{h_{i,j}^{l,u,l^{\prime}}}
		=\sum_{m=1}^{|\mathcal A|} \overleftarrow{p}^{l,u,l^{\prime}}_{m,i,j} \mathcal{CN}(h_{i,j}^{l,u,l^{\prime}}a_m\mid \hat{r}_{i,j}^{w^{l,u,l^{\prime}}}, \tau_{i,j}^{w^{l,u,l^{\prime}}}), 
	\end{align}
	where $\overleftarrow{p}^{l,u,l^{\prime}}_{m,i,j}$ is initialized as $1/|\mathcal A|$ and updated later. Then, given the conditional PDF of $h_{i,j}^{l,u,l^{\prime}}$ in (\ref{hprior}), the message from check nodes $\mathbf e^l$ to variable nodes $\mathbf S$ is the Bernoulli distribution given by
	\begin{align}
	\label{initialS}
		\Delta^{e_{i,j}^{l,u,l^{\prime}}}_{s_{i,j}^{u,l^{\prime}}}
		&\propto \int_{h_{i, j}^{l, u, l^{\prime}}}
		p\left(h_{i, j}^{l, u, l^{\prime}} \mid s_{i, j}^{u, l^{\prime}} \right) \Delta^{g_{i,j}^{l,u,l^{\prime}}}_{h_{i,j}^{l,u,l^{\prime}}}(h_{i,j}^{l,u,l^{\prime}}). 
	\end{align}
    With the inputs $\Delta^{e_{i,j}^{l,u,l^{\prime}}}_{s_{i,j}^{u,l^{\prime}}}(s_{i,j}^{u,l^{\prime}})$, we are now ready to describe the
    messages involved in the MRF. To clearly characterize
    the relative position, the left, right, top, and bottom neighbors of $s_{i,j}^{u,l^{\prime}}$ are reindexed by $\{s_{i,j_{\text L}}^{u,l^{\prime}}, s_{i,j_{\text R}}^{u,l^{\prime}}, s_{i,j_{\text T}}^{u,l^{\prime}}, s_{i,j_{\text B}}^{u,l^{\prime}} \}$. The left, right, top, and bottom input messages of $s_{i,j}^{u,l^{\prime}}$ denoted as $\Omega^{\text{L}^{u,l^{\prime}}}_{i,j}$, $\Omega^{\text{R}^{u,l^{\prime}}}_{i,j}$,$\Omega^{\text{T}^{u,l^{\prime}}}_{i,j}$, and $\Omega^{\text{B}^{u,l^{\prime}}}_{i,j}$, are Bernoulli distribution. The left input message of $s_{i,j}^{u,l^{\prime}}$ can be represented as	
    	$\Omega^{\text{L}^{u,l^{\prime}}}_{i,j} \propto \nonumber\\
    	\int_{\sim s_{i,j}^{u,l^{\prime}}} \prod_{l=0}^{M-1}
    	\Delta^{e_{i,j}^{l,u,l^{\prime}}}_{s_{i,j}^{u,l^{\prime}}} \prod_{p \in \{\text L,\text T,\text B\}} 
    	\Omega^{p^{u,l^{\prime}}}_{i,j_{\text{L}}} 
    	\gamma(s_{i,j_{\text L}}^{u,l^{\prime}}) 
    	\psi(s_{i,j}^{u,l^{\prime}}, s_{i,j_{\text L}}^{u,l^{\prime}})
    $,
	where $\sim s_{i,j}^{u,l^{\prime}}$ represent the variables except $s_{i,j}^{u,l^{\prime}}$. The input messages of $s_{i,j}^{u,l^{\prime}}$ from right, top, and bottom have a similar form to $\Omega^{\text{L}^{u,l^{\prime}}}_{i,j}$. Then, the output message of $s_{i,j}^{u,l^{\prime}}$ is given by
	\begin{align}
	\label{outputS}
		\Delta^{s_{i,j}^{u,l^{\prime}}}_{e_{i,j}^{l,u,l^{\prime}}} \propto 
		\gamma(s_{i,j}^{u,l^{\prime}})
		\prod_{\hat l \neq l} \Delta_{s_{i,j}^{u,l^{\prime}}}^{e_{i,j}^{\hat l,u,l^{\prime}}}
		\prod_{p \in \{\text L, \text R, \text T, \text B\}}\Omega^{\text{p}^{u,l^{\prime}}}_{i,j}.		
	\end{align}	
	The refined messages of $\mathbf H$ after exploring the 3D block sparsity will be fed back to the CCESD module given by
	\begin{align}
	\label{TSEend}
		\Delta^{e_{i,j}^{l,u,l^{\prime}}}_{h_{i,j}^{l,u,l^{\prime}}}
		\propto \int_{s_{i,j}^{u,l^{\prime}}} \Delta^{s_{i,j}^{u,l^{\prime}}}_{e_{i,j}^{l,u,l^{\prime}}}
			p\left(h_{i, j}^{l, u, l^{\prime}} \mid s_{i, j}^{u, l^{\prime}} \right).
	\end{align}
	With $\Delta^{h_{i,j}^{l,u,l^{\prime}}}_{g_{i,j}^{l,u,l^{\prime}}}=	\Delta^{e_{i,j}^{l,u,l^{\prime}}}_{h_{i,j}^{l,u,l^{\prime}}}$, the messages from $\mathbf g$ to $\mathbf t$ is given by
	\begin{align}
	\label{tstart}
		\Delta^{g_{i,j}^{l,u,l^{\prime}}}_{t_{(l-l^{\prime})_M+uM}}
		&\propto \int_{h_{i,j}^{l,u,l^{\prime}},w_{i,j}^{l,u,l^{\prime}}}   \Delta^{h_{i,j}^{l,u,l^{\prime}}}_{g_{i,j}^{l,u,l^{\prime}}}
		\Delta_{g_{i,j}^{l,u,l^{\prime}}}^{w_{i,j}^{l,u,l^{\prime}}} \nonumber \\ 
		&\quad\quad\quad\delta(w_{i,j}^{l,u,l^{\prime}}-h_{i,j}^{l,u,l^{\prime}}t_{(l-l^{\prime})_M+uM})\nonumber \\
		&=\sum_{m=1}^{|\mathcal{A}|} 
		\overrightarrow{p}^{l,u,l^{\prime}}_{m,i,j} \delta(t_{(l-l^{\prime})_M+uM}-a_m).
	\end{align}
	Given the input messages of $t_{(l-l^{\prime})_M+uM}$, the output message of it is given by
	\begin{align}
	\label{tend}
	\Delta^{t_{(l-l^{\prime})_M+uM}}_{g_{i,j}^{l,u,l^{\prime}}}
	= \sum_{m=1}^{|\mathcal{A}|}
	\overleftarrow{p}^{l,u,l^{\prime}}_{m,i,j} \delta(t_{(l-l^{\prime})_M+uM}-a_m),
	\end{align}
	where $\overleftarrow{p}^{l,u,l^{\prime}}_{m,i,j}$ is updated by the product of all the probability related to $\Delta^{g_{i,j}^{l,u,l^{\prime}}}_{t_{(l-l^{\prime})_M+uM}}$ except for $\overrightarrow{p}^{l,u,l^{\prime}}_{m,i,j}$.
 Next, given $\Delta^{h_{i,j}^{l,u,l^{\prime}}}_{g_{i,j}^{l,u,l^{\prime}}}$ and $\Delta^{t_{(l-l^{\prime})_M+uM}}_{g_{i,j}^{l,u,l^{\prime}}}$, the message of feedback from CCESD module to DDAI is the Bernoulli GM distribution given by
	\begin{align}
		\Delta_{w_{i,j}^{l,u,l^{\prime}}}^{g_{i,j}^{l,u,l^{\prime}}}
		\propto
		\int_{\sim w_{i,j}^{l,u,l^{\prime}}} &\Delta^{h_{i,j}^{l,u,l^{\prime}}}_{g_{i,j}^{l,u,l^{\prime}}}
		\Delta^{t_{(l-l^{\prime})_M+uM}}_{g_{i,j}^{l,u,l^{\prime}}} 
		\nonumber \\
		&\delta(w_{i,j}^{l,u,l^{\prime}}-h_{i,j}^{l,u,l^{\prime}}t_{(l-l^{\prime})_M+uM}).
	\end{align}
	Now, the posterior distribution of $w_{i,j}^{l,u,l^{\prime}}$ is approximated as a Bernoulli GM distribution by combing all the input messages of it given by
	\begin{align}
	\label{poststart}
		\Delta_{w_{i,j}^{l,u,l^{\prime}}}
		&\propto
		\Delta_{w_{i,j}^{l,u,l^{\prime}}}^{g_{i,j}^{l,u,l^{\prime}} } \Delta_{g_{i,j}^{l,u,l^{\prime}}}^{w_{i,j}^{l,u,l^{\prime}}}.
	\end{align}
	Similarly, the posterior distribution of $h_{i,j}^{l,u,l^{\prime}}$ is also approximated as a Bernoulli GM distribution given by
	\begin{align}
		\label{posth}
		\Delta_{h_{i,j}^{l,u,l^{\prime}}}
		&\propto
		\Delta_{h_{i,j}^{l,u,l^{\prime}}}^{g_{i,j}^{l,u,l^{\prime}}} \Delta_{g_{i,j}^{l,u,l^{\prime}}}^{h_{i,j}^{l,u,l^{\prime}}}.
	\end{align}
	We can also get the approximated posterior distribution of information symbols as
	\begin{align}
		\label{postend}
		\Delta_{t_{(l-l^{\prime})_M+uM}} = 
		 \sum_{m=1}^{|\mathcal{A}|} 
		 p^{l,u,l^{\prime}}_{m} \delta(t_{(l-l^{\prime})_M+uM}-a_m),	
	\end{align}
	where $p^{l,u,l^{\prime}}_{m} \propto
		\overrightarrow{p}^{l,u,l^{\prime}}_{m,i,j}	\overleftarrow{p}^{l,u,l^{\prime}}_{m,i,j}$.
	Then the mean and variance of $w_{i,j}^{l,u,l^{\prime}}$ with respect to the approximated posterior distribution will be inputted to the GAMP for next iteration. 
	After the algorithm converges, the estimation of $h_{i,j}^{l,u,l^{\prime}}$ is given by its mean with respect to (\ref{posth}), i.e.,
		$\hat{h}_{i,j}^{l,u,l^{\prime}}=\mathrm{E}[h_{i,j}^{l,u,l^{\prime}}|\mathbf Y]$.
	Based on (\ref{postend}), we perform symbol-by-symbol MAP estimation for information symbols as
	\begin{align}
	\label{tE}
	\hat t_{(l-l^{\prime})_M+uM}=\underset{a \in \mathcal A}{\text{arg max}} \quad \Delta_{t_{(l-l^{\prime})_M+uM}}(a).
	\end{align}
	\subsection{Learning the Hyperparameters}
	We now adopt EM algorithm to learn the prior parameters $\mathbf q\triangleq\{\sigma^2, [\bm{\omega}_u]_{u=0}^{U-1},  [\bm{\mu}_u]_{u=0}^{U-1}, [\bm{\phi}_u]_{u=0}^{U-1} \}$. The
	EM algorithm is an iterative technique that increases a lower
	bound on the likelihood at each iteration, and in our case, the hyperparameters are updated in the $i$-th iteration as
	\begin{align}
	\label{em1}
		&(\sigma^2)^{i+1} = \arg \max \limits_{\sigma^2} 
		\mathrm{E}[\log p\left(\mathbf R, \mathbf Y \mid \sigma^2 \right) \mid \mathbf Y, \mathbf q^i],\\
		&(\mu_k^u)^{i+1} = \arg \max \limits_{\mu_k^u}
		\mathrm{E}[\log p\left(\mathbf H, \mathbf S, \mathbf Y \mid \mu_k^u \right) \mid \mathbf Y, \mathbf q^i],\\
		&(\phi_k^u)^{i+1} = \arg \max \limits_{\phi_k^u}
		\mathrm{E}[\log p\left(\mathbf H, \mathbf S, \mathbf Y \mid \phi_k^u \right) \mid \mathbf Y, \mathbf q^i],\\
		&(\omega_k^u)^{i+1} = \arg \max \limits_{\omega_k^u}
		\mathrm{E}[\log p\left(\mathbf H, \mathbf S, \mathbf Y \mid \omega_k^u \right) \mid \mathbf Y, \mathbf q^i] \label{em4},
	\end{align}
	where $\mathbf R$, $\mathbf H$, and $\mathbf S$ are treated as hidden variables, the expectation is with respect to the  posterior distribution approximated by the aforementioned message passing-type algorithm, and we denote $\mathbf R = [\mathbf R^0, \dots, \mathbf R^{M-1}]$ with $\mathbf R^l=\mathbf C^l \mathbf W^l$. By examining the first derivative of the objective function with respect to the variables in (\ref{em1})-(\ref{em4}), we can get the updates of these hyperparameters. The detailed derivation is omitted here for limited spacing.

	The MRF-GM-AMP algorithm can be summarized as follows: Firstly, the DDAI module adopts the GAMP algorithm to output the message of $\mathbf W$ in (\ref{gampout}) given $\mathbf Y$ and $\mathbf C$. Then, the CCESD module generates the initial message of $\mathbf S$ using (\ref{CCESDfirst}) and (\ref{initialS}) based on the output message by DDAI. Then, the TSE module updates the message of $\mathbf H$ using (\ref{outputS}) and (\ref{TSEend}) to exploit its 3D block sparsity and feeds back the refined message to the CCESD module. At the same time, the message of $\mathbf t$ is updated using (\ref{tstart}) and (\ref{tend}). Then, the posterior distribution for $(\mathbf W, \mathbf H, \mathbf t)$ is computed using (\ref{poststart})-(\ref{postend}), and the estimated distribution of $\mathbf W$ is used for the next iteration of the GAMP algorithm. Next, the hyperparameters are updated using (\ref{em1})-(\ref{em4}). The above steps are repeated until convergence. Finally, information symbols and the estimated channel are given by (\ref{tE}) and the expectation with respect to (\ref{posth}), respectively, and the active devices can be detected according to (\ref{detectA}). The complexity of the algorithm is mainly from the GAMP algorithm and the message passing updates, with a total complexity of $\mathcal{O}(QUMN^2N_a + |\mathcal A|KUMNN_a)$, which is linear in the number of users and thus suitable for random access in LEO satellite communications.
	\section{Numerical Results}
	\begin{figure*}
		\centering
		\captionsetup{font={small}}
		\setlength{\belowcaptionskip}{-.6cm}
		\subfigure[Performance of device identification.]{
			\label{AER}
			\begin{minipage}{5.7cm}
				\includegraphics[width=\textwidth]{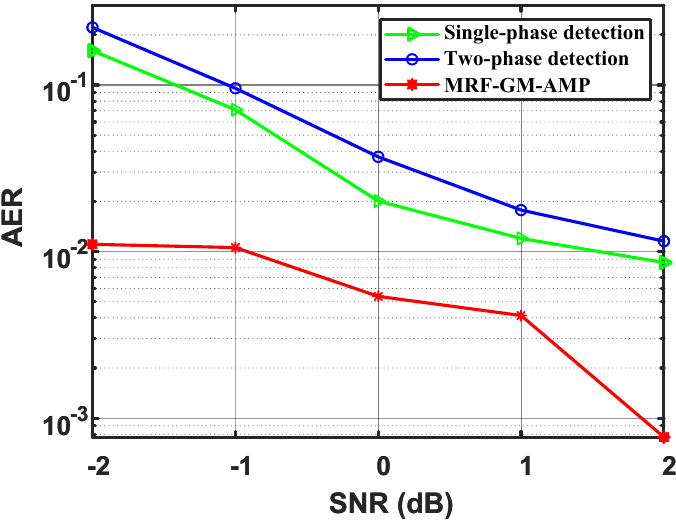} \\
				
			\end{minipage}
		}
		\subfigure[Performance of channel estimation.]{
			\label{NMSE}
			\begin{minipage}{5.7cm}
				\includegraphics[width=\textwidth]{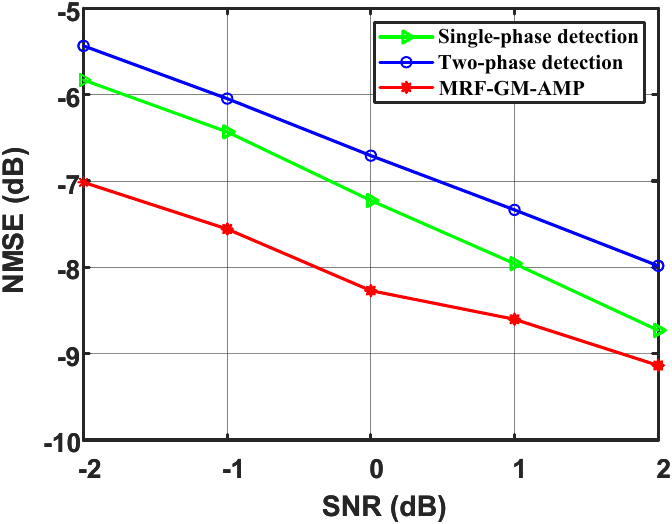} \\
				
			\end{minipage}
		}
		\subfigure[Performance of signal detection.]{
			\label{SER}
			\begin{minipage}{5.7cm} 
				\includegraphics[width=\textwidth]{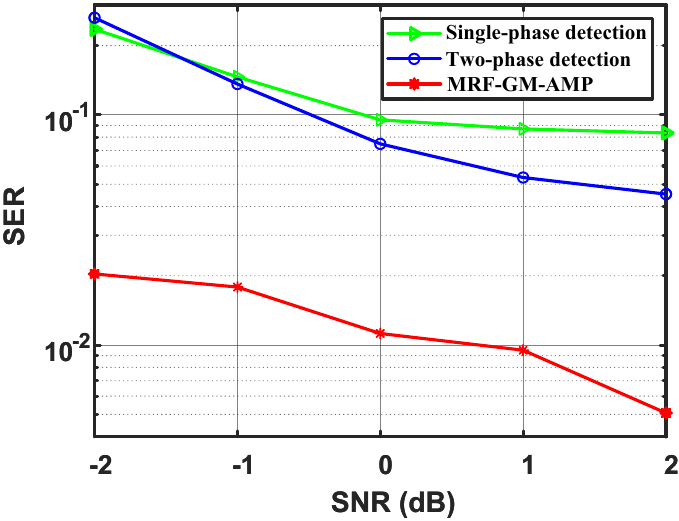} \\
			\end{minipage}
		}
		\caption{Performance comparison between the single-phase detection, two-phase detection, and MRF-GM-AMP under different SNR values, where $U = 40$, $p_{\lambda}=0.1$, and $N_y=N_z=4$.} 
		\label{PCall}
	\end{figure*}
	In this section, we demonstrate the performance of the proposed algorithms through computer simulations. We consider the scenarios of the non-terrestrial networks recommended by the 3GPP\cite{3gpp}, where the satellite operates at S band with $15$ kHz of subcarrier spacing and 600 km of altitude, the power delay profile of channel complex gain follows the NTN-TDL-D, the differential delay (ms) is uniformly selected from $[0,4.44]$, and the Doppler shift (kHz) is uniformly selected from $[-41,41]$. We consider the sporadic transmissions that there are 40 potential devices with 0.1 of active probability, and the active devices transmit consecutive OTFS frames with $Q=20$, $M=16$, and $N=5$. In addition, the nonnegative pulse amplitude
	modulation \cite{Joint2} with $|\mathcal A|=4$ is adopted for solving phase ambiguity, and the elements of spreading code obey $\mathcal{CN}(0,\frac{1}{Q N})$. Finally, we define the received signal-to-noise ratio as $\text{ SNR }=10\log_{10}\frac{\sum_{l=0}^{M-1}\Vert \mathbf R^l\Vert^2_F}{QMNN_a\sigma^2}$, and the average device activity error rate (AER), the average normalized mean-squared-error (NMSE), and the average symbol error rate (SER) are adopted as metrics for device identification, channel estimation, and signal detection, respectively, given by
	$\text{AER} = \frac{1}{U}\sum_{u=0}^{U-1}\left|\lambda_u-\hat \lambda_u\right|$,
	$\text{NMSE}=\frac{\sum_{l=0}^{M-1}\left\| \mathbf H^l-\hat{\mathbf H}^l\right\|^2_{\mathrm{F}}}{\sum_{l=0}^{M-1}\left\|\mathbf H^l\right\|^2_{\mathrm{F}}}$, and
	$\text{SER} = \frac{1}{UM}\sum_{i=0}^{UM-1} \left|t_i - \hat t_i\right|$,
	where we assume that the inactive devices transmit zero for computing the SER.
	
	 Fig. \ref{AER}, Fig. \ref{NMSE}, and Fig. \ref{SER} compare device identification, channel estimation, and signal detection performance between the proposed algorithm and benchmarks, respectively. Here, the single-phase method adopts the same transmission scheme as ours, and the ConvSBL-GAMP \cite{smm1} is used to estimate $\mathbf W^l$ firstly, and then an energy detector is adopted to detect the transmitted information symbols; the two-phase scheme adopts the ConvSBL-GAMP to jointly estimate channel and detect active devices based on transmitted pilots, and then the GAMP detector is adopted to detect information symbols. From the figures, the performance of the proposed algorithm increases with the SNR, and always outperforms the two benchmarks, which indicates the effectiveness of the proposed scheme for LEO satellite-based uplink transmission in presence of the large differential delay and Doppler shift. Notice
	that conventional separated detection scheme has a high
	error floor for SER; hence it can only support a small number of devices. On the other hand, the proposed MRF-GM-AMP works well. This is due to the benefit of joint
	device identification, channel estimation, and signal detection
	design.
	For example, when the AER is around 0.01 in Fig. \ref{AER}, the proposed algorithm has about 3 dB gain in terms of SNR, and in Fig. \ref{NMSE}, the
	proposed algorithm always has 1 dB and 2 dB gain compared with the single-phase and two-phase detection, respectively. In Fig. \ref{SER}, when the SER is around 0.05, the proposed algorithm outperforms the two benchmarks by more than 4 dB. In addition, the SER of MRF-GM-AMP is below 0.05 when the SNR is greater than -2 dB, which indicates that the proposed algorithm could work well in the low SNR regime, and thus is suitable for the satellite communications.
	
	\section{Conclusion}
	This work developed a joint device identification, channel estimation, and signal detection scheme for MIMO-OTFS-based GFRA in LEO satellite communications, where both the large differential delay and Doppler shift exist. To provide low-complexity yet near-optimal estimation and exploit the 3D-structured sparsity of the channel in the delay-Doppler-angle domain, we proposed a message passing-type approach with MRF prior and carefully designed receiver structure. Simulation results demonstrate that the proposed algorithm outperforms conventional algorithms significantly, with a linear complexity in the number of devices and the ability to operate in the low SNR regime, making it suitable for random access in LEO satellite communications.

	\bibliographystyle{IEEEtran}%
	\bibliography{bibfile}
	\vspace{12pt}

\end{document}